\documentclass[prl,twocolumn,amsmath,amssymb,showpacs]{revtex4}
\usepackage{graphicx,bm}

\DeclareMathOperator\sgn{sgn}
\DeclareMathOperator\sinc{sinc}
\renewcommand\Re{\operatorname{Re}}
\renewcommand\Im{\operatorname{Im}}
\newcommand{\ket}[1]{\lvert#1\rangle}
\newcommand\bk{\bm{k}}
\newcommand\bq{\bm{q}}
\newcommand\br{\bm{r}}

\begin{document}

\title{Fermi polarons in two dimensions}

\author{Richard Schmidt}
\author{Tilman Enss}
\affiliation{Physik Department, Technische Universit\"at M\"unchen,
  D-85747 Garching, Germany}
\author{Ville Pietil\"a}
\author{Eugene Demler}
\affiliation{Physics Department, Harvard University,
  Cambridge, Massachusetts 02138, USA}

\date{\today}

\begin{abstract}
  We theoretically analyze inverse radiofrequency (rf) spectroscopy
  experiments in two-component Fermi gases.  We consider a small
  number of impurity atoms interacting strongly with a bath of
  majority atoms.  In two-dimensional geometries we find that the main
  features of the rf spectrum correspond to an attractive polaron and
  a metastable repulsive polaron.  Our results suggest that the
  attractive polaron has been observed in a recent experiment [Phys.\
  Rev.\ Lett.\ \textbf{106}, 105301 (2011)].
\end{abstract}


\maketitle

The behavior of a mobile impurity (polaron) interacting strongly with
a bath of particles is one of the basic many-body problems studied in
condensed matter physics \cite{anderson1967, mitra1987, devreese2009,
  radzihovsky2010}. With the advent of ultracold atomic gases
\cite{bloch2008}, the Fermi polaron problem in which a single
spin-$\downarrow$ atom interacts strongly with a Fermi sea of
spin-$\uparrow$ atoms, has become a subject of intensive research
\cite{chevy2010}. In three dimensions it was found that the polaron
state splits into two branches, a low-energy state interacting
attractively with the bath of fermions, and the repulsive polaron,
which is an excited, metastable state \cite{cui2010, schmidt2011,
  massignan2011}. In this way the polaron exemplifies a more general
paradigm of a many-body system driven into a nonequilibrium state
where a small number of high energy excitations interact strongly with
the surrounding degrees of freedom \cite{schmitt2008, freericks2009}.
The polaron is the limiting case of a Fermi gas with strong spin
imbalance, and the repulsive polaron provides insight into the
question whether a quenched, repulsive Fermi gas may undergo a
transition to a ferromagnetic state even though it is highly excited
\cite{jo2009, cui2010, schmidt2011, massignan2011, pekker2011}.
Similarly, the ground state of the polaron problem has important
implications for the phase diagram of a strongly interacting Fermi gas
\cite{schirotzek2009, nascimbene2009, punk2009}.

\begin{figure}[!ht]
  \centering
  \includegraphics[width=\linewidth]{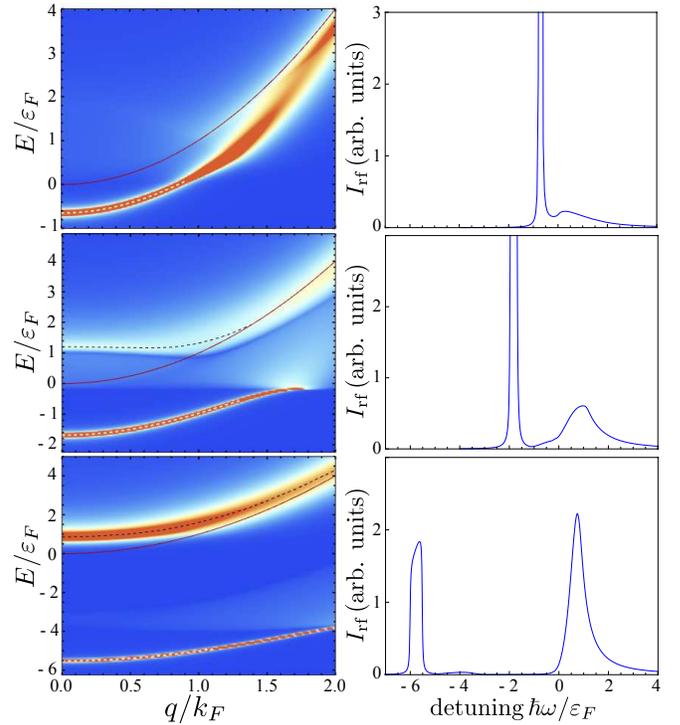}
  \caption{\label{fig:rf} (color online).
    \textbf{Left panels:} spectral function $\mathcal
    A_\downarrow(q,E-\mu_\downarrow)$ for 
    impurity atoms interacting with a 2D Fermi sea.  Red lines
    indicate the free particle dispersion and white (black) dashed
    lines mark the dispersion of the attractive (repulsive) polaron.
    \textbf{Right panels:} corresponding rf spectra illustrating how
    weight is shifted from the attractive polaron state (peak at
    negative frequencies) to the new repulsive polaron state at
    positive frequencies.  The two-body bound state energy is (a)
    $\varepsilon_B/\varepsilon_F = 0.1$, (b)
    $\varepsilon_B/\varepsilon_F = 1$, (c)
    $\varepsilon_B/\varepsilon_F = 5$.}
\end{figure}

It is a key question how many-body properties are affected by reduced
dimensionality, and the polaron is a case in point.  The combination
of optical lattices and Feshbach resonances \cite{bloch2008} provides
a unique setting to experimentally study strongly interacting low
dimensional systems using ultracold atoms \cite{froehlich2011,
  dyke2011}.  Recent advances in radiofrequency (rf) spectroscopy
afford to measure energy spectra \cite{schirotzek2009} and give access
to excited states as well as full spectral functions using momentum
resolved rf \cite{stewart2008, feld2011}.  So far, only the ground state of the
two-dimensional polaron problem has been investigated theoretically
\cite{zoellner2011, parish2011, klawunn2011} with the focus on a
possible polaron to molecule transition.  This is similar to the 3D
situation where for strong interactions it becomes energetically
favorable for the impurity to form a molecular bound state
\cite{prokofev2008fermi}.  The structure of high energy excitations
and the experimental polaron signatures in rf spectroscopy have
remained open questions which we address in this work.  We derive
the spectral functions of both the molecule and the impurity atom
(Fig.~\ref{fig:rf} left) and find that the impurity state splits into
the attractive and the repulsive branch.  We compute rf spectra for
homogeneous 2D systems (Fig.~\ref{fig:rf} right) as well as for the
experimentally relevant quasi-2D geometries (Fig.~\ref{fig:rftrap}).
Finally, we argue that our calculation provides an alternative
explanation of the recent experiment by Fr\"ohlich et
al.~\cite{froehlich2011} in terms of the polaron picture.

A quasi-2D geometry can be realized experimentally using an optical
lattice in one direction with associated trapping frequency
$\omega_z$.  In this case, a confinement induced two-body bound state
exists for an arbitrarily weak attractive interaction
\cite{landau1981, randeria1989, petrov2001} with binding energy
$\varepsilon_B > 0$. The spatial extent of the bound state is related
to the 2D scattering length given by $a_\text{2D}
=\hbar/\sqrt{m\varepsilon_B} > 0$.  In the weak coupling BCS regime of
small 3D scattering length $a_\text{3D} < 0$ \cite{bloch2008} these
dimers are large and weakly bound ($\varepsilon_B \ll \hbar\omega_z$);
in the BEC limit of small $a_\text{3D} > 0$, the weakly interacting
molecules are too tightly bound to feel the confinement
($\varepsilon_B \sim \hbar^2/ (ma_\text{3D}^2) \gg \hbar\omega_z$).
Around the Feshbach resonance ($a_\text{3D}^{-1} = 0$) there is a
strong coupling regime where the binding energy attains the universal
value $\varepsilon_B = 0.244\, \hbar\omega_z$ \cite{petrov2001,
  bloch2008}.

At finite densities the majority atoms form a Fermi gas with Fermi
energy $\varepsilon_F$ and the two-body scattering is replaced by
many-body scattering which gives rise to important qualitative
differences, most notably the emergence of two polaron branches.
Spectral weight is shifted from the attractive to the repulsive
polaron in the non-perturbative regime where the interaction parameter
$1/\ln(\varepsilon_B/ 2\varepsilon_F)$ diverges \cite{bloom1975} and
the confinement induced resonance appears \cite{petrov2000,
  bloch2008}.

We consider a two-component 2D Fermi gas in the limit of extreme spin
imbalance, described by the grand canonical Hamiltonian
\begin{equation*}
  H = \sum_{\bk\sigma} (\varepsilon_{\bk\sigma} - \mu_\sigma)
  c_{\bk\sigma}^\dagger c_{\bk\sigma} + \frac{g}{A} \sum_{\bk\bk'\bq}
  c_{\bk\uparrow}^\dagger c_{\bk'\downarrow}^\dagger
  c_{\bk'-\bq\downarrow} c_{\bk+\bq\uparrow},
\end{equation*}
with single-particle energies $\varepsilon_{\bk\sigma}
=\bk^2/2m_\sigma$ for species $\sigma$ ($\hbar=1$), chemical
potentials $\mu_\sigma$ and system area $A$.  Having in mind the
experiment of Ref.~\cite{froehlich2011}, we focus on the case of equal
masses $m_\uparrow = m_\downarrow = m$. Generalizations to mass
imbalanced situations are straightforward \cite{zoellner2011,
  parish2011}.  In the low-energy limit the attractive $s$-wave
contact interaction $g$ can act only between different species due to
the Pauli principle.  The majority atoms are not renormalized by the
presence of a single impurity with finite mass such that $\mu_\uparrow
= \varepsilon_F=k_F^2/2m$ at zero temperature.  The chemical potential
$\mu_\downarrow$ of the impurity atom is determined such that the
impurity state $\ket\downarrow$ has vanishing macroscopic occupation.
Furthermore, $\mu_\downarrow$ is negative due to the attractive
interaction between $\uparrow$ and $\downarrow$ atoms.

\paragraph{Dressed molecule.}
The two-body scattering of a spin-$\uparrow$ atom and a
spin-$\downarrow$ atom is described by the exact two-body $T$-matrix
\cite{randeria1989}
\begin{equation}
  \label{eq:T0}
  T_0(E) = \frac{4\pi/m}{\ln(\varepsilon_B/E) + i\pi}\;.
\end{equation}
The pole of the $T$-matrix at $E=-\varepsilon_B$ corresponds to the
molecular bound state, and the associated vacuum scattering amplitude
for two particles with relative momenta $\bk$ and $-\bk$ in the
center-of-mass frame is $f(k=|\mathbf k|) = mT_0(2\varepsilon_{\bk}) =
4\pi/[\ln(1/k^2a_\text{2D}^2)+i\pi]$ \cite{bloch2008}.

\begin{figure}[t]
  \centering
  \includegraphics[width=\linewidth]{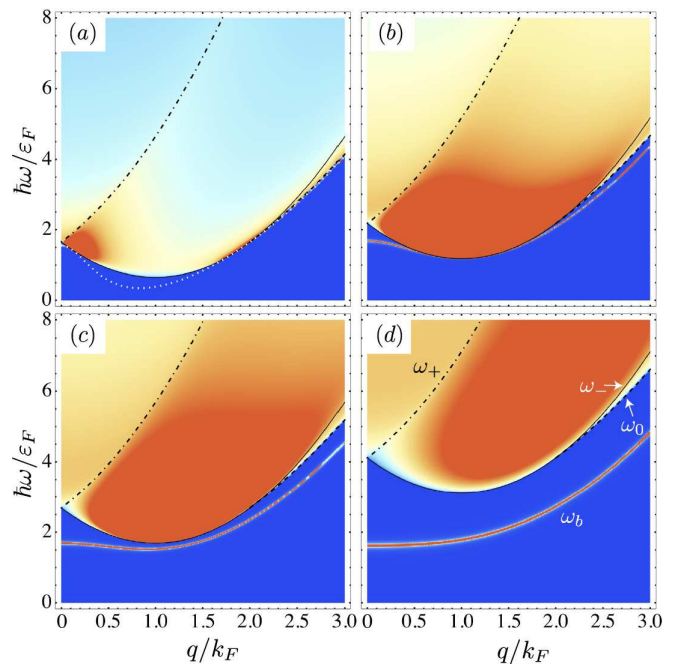}
  \caption{\label{fig:mol} (color online).  Molecular spectral
    function $\mathcal{A}_\text{mol}(\bq,\omega)$ for different values
    of the two-body binding energy $\varepsilon_B/\varepsilon_F$: (a) $0.1$,
    (b) $0.5$, (c) $1.0$, (d) $2.5$.  The dashed lines mark the log
    continuum $\omega_0$, dash-dotted and solid the root continuum
    $\omega_\pm$.}
\end{figure}

In the presence of a Fermi sea of spin-$\uparrow$ atoms, the molecular
state is dressed by fluctuations and described by the many-body
$T$-matrix.  This can be calculated in the Nozieres--Schmitt-Rink
approach \cite{nozieres1985}, as done in the 2D case by Engelbrecht
and Randeria \cite{engelbrecht1990, engelbrecht1992}.  We generalize
these results to the case of spin imbalance and obtain
\begin{multline}
  \label{eq:Tinv}
  T^{-1}(\bq,\omega) =
  T_0^{-1}(\omega+i0+\mu_\uparrow+ \mu_\downarrow - \varepsilon_{\bq}/2) \\
  + \int \frac{d^2k}{(2\pi)^2}\,
  \frac{n_F(\varepsilon_{\bk}-\mu_\uparrow) +
    n_F(\varepsilon_{\bk+\bq}-\mu_\downarrow)}
  {\omega + i0 + \mu_\uparrow + \mu_\downarrow - \varepsilon_{\bk}
    - \varepsilon_{\bk+\bq}},
\end{multline}
with the Fermi function $n_F(\varepsilon)$.  At zero temperature where
$\mu_\uparrow = \varepsilon_F$ and $\mu_\downarrow < 0$, we obtain an
analytical expression for the many-body $T$-matrix
\begin{equation}
  \label{eq:E}
  T(\bq,\omega) = T_0\Bigl(\tfrac 12 z \pm \tfrac 12
  \sqrt{(z-\varepsilon_{\bq})^2-4\varepsilon_F
    \varepsilon_{\bq}} \Bigr)
\end{equation}
with $z=\omega+i0 - \varepsilon_F + \mu_\downarrow$ and $\pm = \sgn
\Re (z-\varepsilon_{\bq})$.  Due to the constant density of states in
2D the many-body $T$-matrix can be expressed as the two-body
$T$-matrix with the argument shifted by Pauli blocking.  The molecular
spectral function $\mathcal A_\text{mol}(\bq,\omega) = -2\Im
T(\bq,\omega)$ is shown in Fig.~\ref{fig:mol} for several values of
the interaction strength parametrized by the two-body binding energy
$\varepsilon_B$.  One observes a bound state peak at low energies and
the particle-particle continuum at higher energies.

The continuum of dissociated molecules arises mathematically from the
branch cut of the square root \eqref{eq:E} in the region $\omega_-(q)
< \omega < \omega_+(q)$, $\omega_\pm = \varepsilon_F (1\pm q/k_F)^2 -
\mu_\downarrow$ (dash-dotted/solid lines), as well as from the branch
cut of the logarithm \eqref{eq:T0} for $\omega > \omega_+(q)$ and for
$\omega_0=\varepsilon_{\bq}/2-\varepsilon_F - \mu_\downarrow < \omega
< \omega_-(q)$ if $q>2k_F$ (dashed lines).

The bound state pole of the many-body T-matrix has the dispersion
relation \cite{zoellner2011}
\begin{align}
  \label{eq:molbnd}
  \omega_b(\bq) =
  \frac{\varepsilon_{\bq}/2(\varepsilon_{\bq}/2-\varepsilon_F) +
    \varepsilon_B (\varepsilon_F-\varepsilon_B)} {\varepsilon_{\bq}/2 +
    \varepsilon_B} - \mu_\downarrow
\end{align}
which changes qualitatively with the two-body binding energy
$\varepsilon_B$ (see Fig.~\ref{fig:mol}): for $\varepsilon_B <
2\varepsilon_F$ the bound state has minimum energy at a finite wave
vector with positive effective mass $m^*/m =
(2-2/k_Fa_\text{2D})^{-1}$ \cite{suppl_mat}.

\begin{figure}[!ht]
  \centering
  \includegraphics[width=\linewidth]{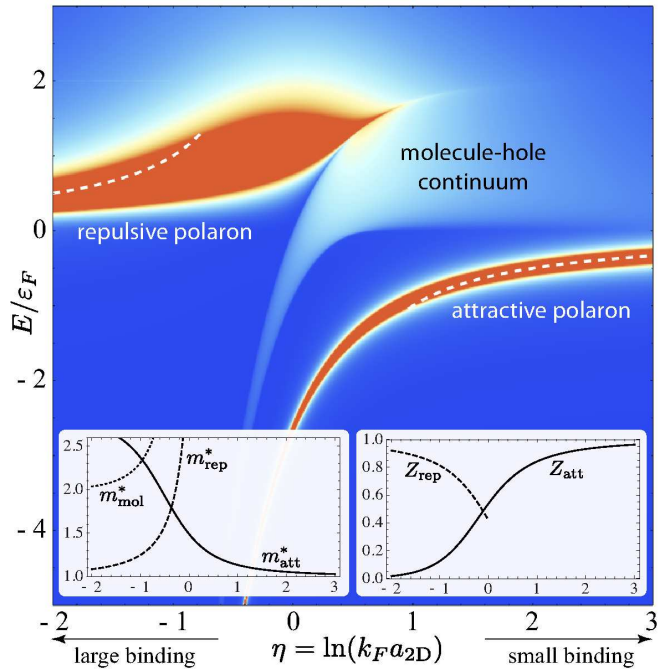}
  \caption{\label{fig:spec} (color online).  Polaron spectral function
    $\mathcal{A}_\downarrow(q=0,E-\mu_\downarrow)$ versus the interaction
    parameter $\eta$.  The dashed lines indicate the perturbation theory of 
    Ref.~\cite{bloom1975}.  \textbf{Left inset:} effective mass $m^*/m$ of
    the attractive and repulsive polaron as well as the molecule.
    \textbf{Right inset:} crossover of the quasi-particle weight
    $Z$ from the repulsive to the attractive polaron.}
\end{figure}

\begin{figure}[ht]
  \centering
  \includegraphics[width=\linewidth]{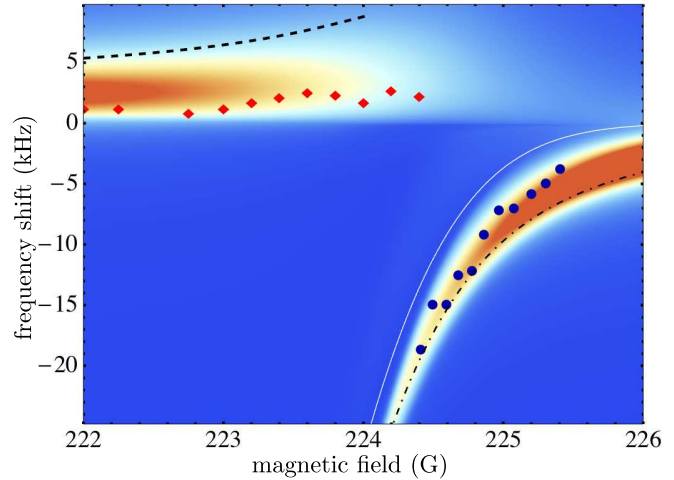}
  \caption{\label{fig:rftrap} (color online).  Trap averaged rf
    spectra of a quasi-2D Fermi gas: rf detuning versus magnetic field
    B. The experimental data points (blue and red diamonds) are taken
    from Ref.~\cite{froehlich2011}. Also shown are the energy of the
    repulsive (dashed) and attractive (dash-dotted) polaron as well as
    the two-body binding energy (solid, white) in a homogeneous
    system.}
\end{figure}

\paragraph{Polaron and quasi-particle properties.}
The impurity atom is dressed with virtual molecule-hole excitations
and becomes a quasi-particle with self-energy \cite{engelbrecht1992,
  punk2007, suppl_mat}
\begin{equation}
  \label{eq:Sdown}
  \Sigma_\downarrow(\bq,\omega) = \int_{k<k_F} \frac{d^2k}{(2\pi)^2}\,
  T(\bk+\bq,\varepsilon_{\bk}-\mu_\uparrow + \omega) \,,
\end{equation}
which leads to the same ground state energy as a variational ansatz
\cite{combescot2007}.  Here we have used the fact that in the
zero-temperature polaron problem the molecule has vanishing
macroscopic occupation. Hence it has spectral weight only at positive
frequencies (cf.\ Fig.~\ref{fig:mol}) where the Bose distribution
vanishes.  We perform the integral in \eqref{eq:Sdown} numerically and
obtain the spectral function of impurity atoms
\begin{equation}
  \label{eq:Adown}
  \mathcal A_\downarrow(\bq,\omega) = -2 \Im [\omega+i0+\mu_\downarrow
  - \varepsilon_{\bq} - \Sigma_\downarrow(\bq,\omega)]^{-1}.
\end{equation}
The frequency and momentum dependence of the spectral function is
shown in Fig.~\ref{fig:rf} (left panel) for three values of the
interaction strength. In Fig.~\ref{fig:spec} we display the
zero-momentum spectral function $\mathcal
A_\downarrow(q=0,E-\mu_\downarrow)$ versus interaction parameter $\eta
= \ln(k_Fa_\text{2D}) = -\ln(\varepsilon_B/ 2\varepsilon_F)/2$.  In
both figures we set the reference energy to the free atom threshold by
subtracting the chemical potential $\mu_\downarrow$.

At weak binding $\varepsilon_B\ll \varepsilon_F$ (Fig.~\ref{fig:rf}a)
the attractive polaron is a well-defined quasi-particle at small
momenta but for $q\gtrsim k_F$ it scatters off virtual molecules and
acquires a large decay width.  For intermediate binding
(Fig.~\ref{fig:rf}b) a new repulsive polaron state appears at positive
energies.  It is a metastable state with broad decay width, and it is
shifted to higher energy due to the repulsive interaction with the
Fermi sea of spin-$\uparrow$ atoms.  The dispersion of the repulsive
polaron has a minimum at finite momentum $q \sim k_F$ reflecting a
similar feature in the molecular spectral function
(Fig.~\ref{fig:mol}c); for larger momenta it approaches the free
particle dispersion.  Finally, for strong binding (Fig.~\ref{fig:rf}c)
both polaron branches are well separated and the repulsive polaron
becomes an increasingly long-lived and stable quasi-particle.  Between
the attractive and the repulsive polaron branches appears the
molecule-hole continuum (see also Fig.~\ref{fig:spec}).  Its spectral
weight is small in the case of a broad Feshbach resonance studied
here, but it is enhanced for narrow resonances by an admixture of
closed channel molecules \cite{schmidt2011}.

It is instructive to see how the quasi-particle properties of the
polaron change as the interaction parameter $\eta$ is varied.  The
right inset of Fig.~\ref{fig:spec} shows a continuous crossover where
the quasi-particle weight $Z=1/[1-\partial_\omega
\Sigma(\bq=0,\omega)]$
evaluated at the quasi-particle pole, shifts from the attractive to the
repulsive polaron branch: for small binding ($\eta>0$), the attractive
polaron is the dominant excitation and the weight is gradually
transferred toward the repulsive branch for increasing binding
($\eta<0$).  This crossover is also reflected in the effective mass
$m^*/m$ (Fig.~\ref{fig:spec} left inset). Our strong coupling
calculation reproduces the perturbative results \cite{bloom1975} for
the attractive and repulsive polaron energies in the weak and strong
binding limits (dashed lines in Fig.~\ref{fig:spec}).

\paragraph{Radiofrequency spectroscopy.} 
The spectral properties of the imbalanced Fermi gas can be accessed
experimentally using rf spectroscopy.  We assume that an rf pulse is
used to drive atoms from an initial state $\ket{i}$ to an initially
empty final state $\ket{f}$.  We choose the final state to be strongly
interacting with a bath of a third species $\ket\uparrow$ such that
$\ket f$ is in fact the impurity state, $\ket f=\ket\downarrow$.  This
inverse rf procedure interchanges the roles of $\ket i$ and $\ket f$
with respect to Ref.~\cite{schirotzek2009}; it has been proposed in
Refs.~\cite{schmidt2011, massignan2011} and realized in the experiment
by Fr\"ohlich et al.\ \cite{froehlich2011}.

Within linear response theory, the rf transition rate is given by
\cite{haussmann2009}
\begin{equation}
  I_\text{rf}(\omega_\text{rf}) = -2\Omega_\text{rf}^2 
  \Im \chi^R (-\omega_\text{rf}-\mu_i+\mu_f)
\end{equation}
where $\Omega_\text{rf}$ is the Rabi frequency, $\omega_\text{rf}$ the
detuning of the rf photon from the bare transition frequency and
$\mu_{i(f)}$ the initial (final) state chemical potential.  The
retarded correlation function $\chi^R$ can be computed from the
corresponding time-ordered correlation function $-i\theta(t-t')
\langle [\psi_f^\dagger(\br,t) \psi_i(\br,t), \psi_i^\dagger(\br',t')
\psi_f(\br',t')] \rangle$ \cite{punk2007, massignan2008}.  In general
vertex corrections are crucial \cite{pieri2009, enss2011}, but we find
that they vanish in the case of negligible initial state interactions
as appropriate for the experiment \cite{froehlich2011}. At $T=0$, we obtain
\begin{equation}
  \label{eq:rf_current}
  I_\text{rf}(\omega_\text{rf})
  = \Omega_\text{rf}^2 \int_{\varepsilon_{\bq}<\mu_i} \frac{d^2q}{(2\pi)^2}\,
  A_\downarrow(\bq,\omega_\text{rf}+\varepsilon_{\bq}-\mu_\downarrow)\,.
\end{equation} 
The integral in equation \eqref{eq:rf_current} is calculated numerically and the resulting
rf spectra are shown in Fig.~\ref{fig:rf} (right panel).  The rf
probes the final $\ket\downarrow$ state spectral function along the
free-particle dispersion up to the initial state chemical potential
$\mu_i$.  As in the experiment \cite{froehlich2011}, we assume a
balanced initial state mixture with $\mu_i = \mu_\uparrow$.  We find a
peak in the rf spectrum once the detuning $\omega_\text{rf}$ reaches
the final state chemical potential $\mu_\downarrow$ ($\mu_\downarrow$
is negative in the polaron problem).  The transfer of spectral weight
from the attractive to the repulsive polaron can be directly observed
in Fig.~\ref{fig:rf}.

\paragraph{Comparison to experiments.}
In order to relate our results to harmonically confined Fermi gases,
we have to connect the strict 2D calculation to the quasi-2D geometry
relevant to experiments~\cite{petrov2001, bloch2008}.  Well below the
confinement energy $\hbar\omega_z$ where only the lowest transverse
mode is occupied, this can be done by replacing $\varepsilon_B$ with
the exact quasi-2D two-body binding energy. Thus $\varepsilon_B$
becomes a function of both the 3D scattering length $a_\text{3D}$ and
the confinement length $\ell_z =
\sqrt{\hbar/m\omega_z}$~(\cite{petrov2001}, cf. Eq.~82 in
\cite{bloch2008}).

Recently the quasi-2D geometry has been realized experimentally with
a Fermi gas of $^{40}$K atoms \cite{froehlich2011}.  Following the
inverse rf procedure described above, an initially non-interacting
balanced mixture is driven into a strongly interacting final state.
As long as its occupation remains small the final state is a
Fermi polaron, and our calculation predicts the experimental rf
response.

In Fig.~\ref{fig:rftrap} we show our trap averaged rf spectra versus
magnetic field.  We use the experimental parameters of
Ref. \cite{froehlich2011} with $\omega_z = 2\pi\times 80$~kHz,
$\omega_\perp = 2\pi \times 125$~Hz, and express $a_\text{3D}$ in
terms of the magnetic field \cite{bloch2008,strohmaier2010}. We
incorporate the radial trapping in the 2D plane using the local
density approximation; the local Fermi energy is $\varepsilon_F(r) =
\varepsilon_F - m\omega_\perp^2 r^2/2$ with peak Fermi energy
$\varepsilon_F = 9$~kHz.  Finally, we average over 30 pancakes in the
$z$ direction \cite{froehlich2011}.

We observe that the lower branch of the experimental spectra (circles)
agrees well with the attractive polaron picture
(Fig.~\ref{fig:rftrap}) and our calculation provides an alternative
interpretation to the two-body bound state (solid line) put forward in
Ref.~\cite{froehlich2011}.  We note that also the measured frequency
shift in 3D as shown in~\cite{froehlich2011} fits the polaron
picture~\cite{schmidt2011}. Our results show a second rf peak at
positive detunings corresponding to the repulsive polaron.  The dashed
line in Fig.~4 indicates its quasi-particle energy in the bulk (cf.\
Fig.~\ref{fig:spec}).  As similar for the attractive polaron energy
(dash-dotted line) the trap average leads to a significant shift of
the rf peaks to lower energies.  The experimental data (diamonds) in
this magnetic field range agrees qualitatively with our
calculation. One possible reason for the remaining discrepancy is the
large final state occupation in the experiment.

In conclusion, we studied Fermi polarons in two dimensions which
exhibit an attractive and repulsive branch and computed their rf
spectra. Additional work is needed to understand discrepancies between
theory and experiment for repulsive polarons. As an example, pump and
probe experiments in the form of a sequence of two short pulses may
shed further light on this issue.

\begin{acknowledgments}
  We thank M.~K\"ohl and his group for kindly providing us with their
  data, and M.~K\"ohl, M.~Punk and W.~Zwerger for useful discussions.
  We acknowledge financial support from Harvard-MIT CUA, DARPA OLE
  program, NSF Grant No. DMR-07-05472, AFOSR Quantum Simulation MURI,
  AFOSR MURI on UltracoldMolecules, and the ARO-MURI on Atomtronics
  (VP and ED), Academy of Finland (VP) and DFG through FOR 801 (RS and
  TE).
\end{acknowledgments}



\section*{Supplementary material}

\paragraph{Many-body $T$-matrix.}
The molecular spectral function displayed in Fig.~\ref{fig:mol}
changes qualitatively with the binding energy.  For large binding
$\varepsilon_B > 2\,\varepsilon_F$ (Fig.~\ref{fig:mol}d), the bound
state \eqref{eq:molbnd} has minimum energy $\omega_b = \varepsilon_F -
\varepsilon_B - \mu_\downarrow$ at $q=0$ and we find an effective mass
$m^*/m\rvert_{q=0} = 2/(1-2\varepsilon_F / \varepsilon_B) > 2$.  While
it is well known that the $T$-matrix does not yield the correct
binding energy of the molecule in the BEC limit \cite{parish2011} the
repulsive polaron, which is the focus of our work, is reproduced
correctly \cite{bloom1975}.  For smaller binding $\varepsilon_B <
2\,\varepsilon_F$ (Fig.~\ref{fig:mol}b-c), the effective mass at $q=0$
becomes negative and a new minimum appears at finite wave vector $q_c
= 2\sqrt{k_Fa_\text{2D}-1}/a_\text{2D}$ \cite{zoellner2011} and with
positive effective mass $m^*/m\rvert_{q=q_c} =
(2-2/k_Fa_\text{2D})^{-1}$.  For decreasing binding energy
$\varepsilon_B$ the bound state eventually touches the continuum at
momenta $q_\pm/k_F = 1\pm \sqrt{1-2\varepsilon_B/ \varepsilon_F}$
(dotted line in Fig.~\ref{fig:mol}a).  For $q_- < q < q_+$ the bound
state has negative residue and it ceases to exist.

\begin{figure}[!ht]
  \centering
  \includegraphics[width=0.9\linewidth]{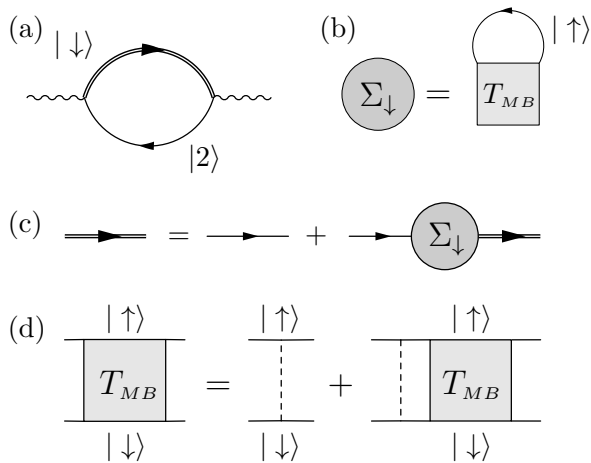}
  \caption{\label{fig:diagrams} Diagrammatic representation of (a) the
    rf photon self-energy, (b) the self-energy of spin-$\downarrow$
    atoms, (c) the Dyson equation for the dressed Green's function
    $G_\downarrow$, and (d) the many-body T-matrix between states
    $\ket\uparrow$ and $\ket\downarrow$.}
\end{figure}

\paragraph{Polaron self-energy.}
We compute the many-body $T$-matrix in the ladder approximation which
is represented diagrammatically in Fig.~\ref{fig:diagrams}d.  The
self-energy of the $\downarrow$ atom is given by scattering an
$\uparrow$ hole off a molecule as depicted in
Fig.~\ref{fig:diagrams}b.  Explicitly, the self-energy reads
\begin{multline}
  \label{eq:sedr}
  \Sigma_\downarrow^R(\bq,\omega)
  = \int\frac{d^2k}{(2\pi)^2} \frac{dz}{\pi}
  \bigl[  n_B(z) G^A_{\uparrow}(\bk-\bq,z-\omega) \Im T(\bk,z) \\ 
  - n_F(z) \Im G^R_{\uparrow}(\bk,z)\, T(\bk+\bq,z+\omega) \bigr]
\end{multline}
where $G^{A,R}_{\uparrow}$ refers to advanced/retarded $\uparrow$
Green's functions.  The first contribution comes from the pole and
branch cuts of the $T$-matrix.  In the polaron problem neither the
$\downarrow$ state nor the molecular state are macroscopically
occupied at zero temperature, hence the molecular spectral function
$\propto \Im T$ has weight only at positive frequencies $z>0$ where
$n_B(z)$ vanishes, cf.\ Fig.~\ref{fig:mol}.  The second contribution
of \eqref{eq:sedr} with the bare $\uparrow$ spectral function $\Im
G^R_\uparrow(\bk,z) = -i\pi \delta(z-\varepsilon_{\bk}+\mu_\uparrow)$
directly yields Eq.~\eqref{eq:Sdown}.  The $\downarrow$ self-energy
yields the $\downarrow$ Green's function via the Dyson equation
\eqref{eq:Adown} depicted diagrammatically in
Fig.~\ref{fig:diagrams}c.

\paragraph{Local density approximation.}
To incorporate the radial trapping potential we use the local density
approximation (LDA). In the experiment \cite{froehlich2011} not only a
single (central) 2D layer is populated but also additional pancakes
along the axial direction. For each of these layers we calculate the
rf response
\begin{equation}
  I_\text{trap}^{(\nu)}(\omega)
  = \int d^2r\, \rho_\text{LDA}^{(\nu)}(\br)\, I_\text{rf}^{(\nu)}(\omega,\br).
\end{equation}
\begin{figure}[b]
  \centering
  \includegraphics[width=.9\linewidth]{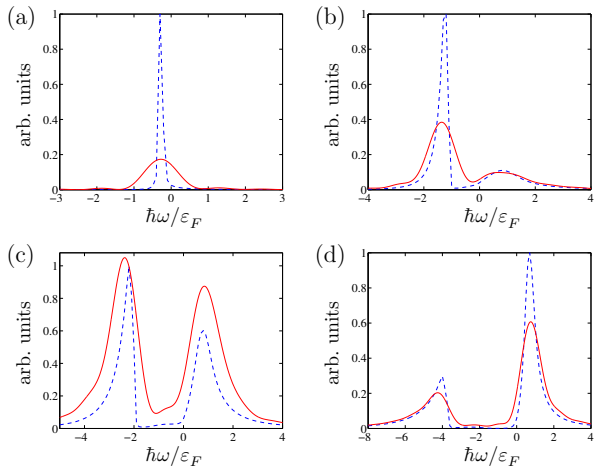}
  \caption{\label{fig:rfconv1} (color online). RF spectrum of a
    trapped quasi-2D Fermi gas for (a) $B=227$ G, (b) $B=225$ G, (c)
    $B=224.5$ G, and (d) $B=224$ G.  The attractive polaron gives rise
    to the peak at negative frequencies and the repulsive polaron
    corresponds to the peak at positive frequencies.  The dashed line
    depicts the trap averaged signal $I_\text{trap}$ while the solid
    line shows the expected experimental signal $I_\text{exp}$ which
    takes into account a rectangular rf pulse of duration $T=100\,\mu$s (not
    to scale) \cite{froehlich2011}.}
\end{figure}

The various pancakes are indicated by the index $\nu$ and
$\rho_\text{LDA}^{(\nu)}(\br) = m\varepsilon_F^{(\nu)}(r)/2\pi\hbar^2$
is the Thomas-Fermi distribution of the density of non-interacting
fermions within a single layer.  $I_\text{rf}^{(\nu)}(\omega,\br)$ is
the rf response in Eq.~\eqref{eq:rf_current} computed for a
homogeneous system with local Fermi energy $\varepsilon_F^{(\nu)}(r)$
and interaction parameter $\eta^{(\nu)}(r) =
-\ln(\varepsilon_B/2\varepsilon_F^{(\nu)}(r))/2$.  In the experiment
\cite{froehlich2011} 30 layers have been populated. To obtain the
complete rf response of the trapped system we finally sum over 30
contributions $I_\text{trap}^{(\nu)}$ where we assume that the peak
density of each layer, $\epsilon_F^{(\nu)}(0)$, varies according to a
Thomas-Fermi profile along the lattice direction.

\begin{figure}[b]
  \centering
  \includegraphics[width=.9\linewidth]{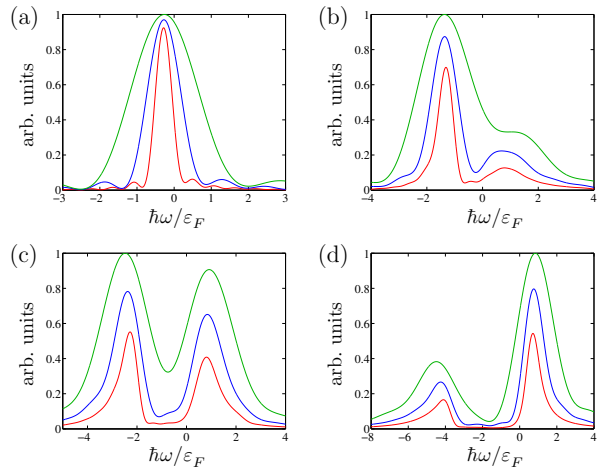}
  \caption{\label{fig:rfconv2} (color online). Same as
    Fig.~\ref{fig:rfconv1} but for different rf pulse duration (from
    top to bottom) $T=50\,\mu$s, $T=100\,\mu$s, and $T=200\,\mu$s (not
    to scale).}
\end{figure}

\paragraph{Shape and length of the rf pulse.}
The line shape of the rf spectra has a strong dependence on the rf
pulse shape.  The rf pulse used in Ref.~\cite{froehlich2011} is
approximately rectangular with duration $T=100\,\mu$s.  We compute the
experimental rf signal as the convolution of $I_\text{trap}(\omega)$
with the Fourier spectrum of the rf field \cite{haussmann2009}
\begin{equation}
  I_\text{exp}(\omega)
  = \frac{T}{2\pi} \int d\omega'\, I_\text{trap}(\omega-\omega')\, \sinc^2(\omega'T/2). 
\end{equation}
In Fig.~\ref{fig:rfconv1} we show the resulting rf spectra for
different values of the external magnetic field.  Note that the
broadening may shift the apparent peak position, see e.g.\
Fig.~\ref{fig:rfconv2}b.


\begin{thebibliography}{38}

\bibitem{anderson1967}
P.~W. Anderson, Phys.\ Rev.\ Lett.\ \textbf{18}, 1049 (1967).

\bibitem{mitra1987}
T.~K. Mitra, A.~Chatterjee, and S.~Mukhopadhyay, Phys.\ Rep.\ \textbf{153}, 91
  (1987).

\bibitem{devreese2009}
J.~T. Devreese and A.~S. Alexandrov, Rep.\ Prog.\ Phys.\ \textbf{72}, 066501
  (2009).

\bibitem{radzihovsky2010}
L.~Radzihovsky and D.~E. Sheehy, Rep.\ Prog.\ Phys.\ \textbf{73}, 076501 (2010).

\bibitem{bloch2008}
I.~Bloch, J.~Dalibard, and W.~Zwerger, Rev.\ Mod.\ Phys.\ \textbf{80}, 885
  (2008).

\bibitem{chevy2010}
F.~Chevy and C.~Mora, Rep.\ Prog.\ Phys.\ \textbf{73}, 112401 (2010).

\bibitem{cui2010}
X.~Cui and H.~Zhai, Phys.\ Rev.~A \textbf{81}, 041602 (2010).

\bibitem{schmidt2011}
R.~Schmidt and T.~Enss, Phys.\ Rev.~A \textbf{83}, 063620 (2011).

\bibitem{massignan2011}
P.~Massignan and G.~M. Bruun, Eur.\ Phys.~J.~D  \textbf{65}, 83 (2011). 

\bibitem{schmitt2008}
F.~Schmitt{ }\textit{et al.}, Science \textbf{321}, 1649
  (2008).

\bibitem{freericks2009}
J.~K. Freericks, H.~R. Krishnamurthy, and T.~Pruschke, Phys.\ Rev.\ Lett.\
  \textbf{102}, 136401 (2009).

\bibitem{jo2009}
G.~B. Jo{ }\textit{et al.}, Science \textbf{325}, 1521 (2009).

\bibitem{pekker2011}
D.~Pekker{ }\textit{et al.}, Phys.\ Rev.\ Lett.\ \textbf{106}, 050402 (2011).

\bibitem{schirotzek2009}
A.~Schirotzek, C.-H. Wu, A.~Sommer, and M.~W. Zwierlein, Phys.\ Rev.\ Lett.\
  \textbf{102}, 230402 (2009).

\bibitem{nascimbene2009}
S.~Nascimb\`ene{ }\textit{et al.}, Phys.\ Rev.\ Lett.\ \textbf{103}, 170402 (2009).

\bibitem{punk2009}
M.~Punk, P.~T. Dumitrescu, and W.~Zwerger, Phys.\ Rev.~A \textbf{80}, 053605
  (2009).

\bibitem{froehlich2011}
B.~Fr{\"o}hlich{ }\textit{et al.},
  Phys.\ Rev.\ Lett.\ \textbf{106}, 105301 (2011).

\bibitem{dyke2011}
P.~Dyke{ }\textit{et al.}, Phys.\ Rev.\ Lett.\ \textbf{106}, 105304
(2011).

\bibitem{stewart2008}
J.~T. Stewart, J.~P. Gaebler, and D.~S. Jin, Nature \textbf{454}, 744 (2008).

\bibitem{feld2011}
M.~Feld, B.~Fr\"ohlich, E.~Vogt, M.~Koschorrek, and M.~K\"ohl, Nature \textbf{480}, 75 (2011).

\bibitem{zoellner2011}
S.~Z{\"o}llner, G.~M. Bruun, and C.~J. Pethick, Phys.\ Rev.~A \textbf{83},
  021603(R) (2011).

\bibitem{parish2011}
M.~M. Parish, Phys.\ Rev.~A \textbf{83}, 051603(R) (2011).

\bibitem{klawunn2011}
M.~Klawunn and A.~Recati, Phys.\ Rev.~A \textbf{84}, 033607 (2011).

\bibitem{prokofev2008fermi}
N.~V. Prokof{'}ev and B.~V. Svistunov, Phys.\ Rev.~B \textbf{77}, 020408
  (2008).

\bibitem{landau1981}
L.~D. Landau and E.~M. Lifshitz, \textit{Quantum Mechanics}
  (Butterworth-Heinemann, Oxford, 1981).

\bibitem{randeria1989}
M.~Randeria, J.-M. Duan, and L.-Y. Shieh, Phys.\ Rev.\ Lett.\ \textbf{62}, 981
  (1989).

\bibitem{petrov2001}
D.~S. Petrov and G.~V. Shlyapnikov, Phys.\ Rev.~A \textbf{64}, 012706 (2001).

\bibitem{bloom1975}
P.~Bloom, Phys.\ Rev.~B \textbf{12}, 125 (1975).

\bibitem{petrov2000}
D.~S. Petrov, M.~Holzmann, and G.~V. Shlyapnikov, Phys.\ Rev.\ Lett.\
  \textbf{84}, 2551 (2000).

\bibitem{nozieres1985}
P.~Nozieres and S.~Schmitt-Rink, J. Low Temp.\ Phys. \textbf{59}, 195 (1985).

\bibitem{engelbrecht1990}
J.~R. Engelbrecht and M.~Randeria, Phys.\ Rev.\ Lett.\ \textbf{65}, 1032 (1990).

\bibitem{engelbrecht1992}
J.~R. Engelbrecht and M.~Randeria, Phys.\ Rev.~B \textbf{45}, 12419 (1992).

\bibitem{suppl_mat}
See Supplementary Material for details of the derivation.

\bibitem{punk2007}
M.~Punk and W.~Zwerger, Phys.\ Rev.\ Lett.\ \textbf{99}, 170404 (2007).

\bibitem{combescot2007}
R.~Combescot, A.~Recati, C.~Lobo, and F.~Chevy, Phys.\ Rev.\ Lett.\
  \textbf{98}, 180402 (2007).

\bibitem{haussmann2009}
R.~Haussmann, M.~Punk, and W.~Zwerger, Phys.\ Rev.~A \textbf{80}, 063612
  (2009).

\bibitem{massignan2008}
P.~Massignan, G.~M. Bruun, and H.~T.~C. Stoof, Phys.\ Rev.~A \textbf{77},
  031601(R) (2008).

\bibitem{pieri2009}
P.~Pieri, A.~Perali, and G.~C. Strinati, Nature Phys.\ \textbf{5}, 736 (2009).

\bibitem{enss2011}
T.~Enss, R.~Haussmann, and W.~Zwerger, Ann.\ Phys.\ (N.Y.) \textbf{326}, 770
  (2011).

\bibitem{strohmaier2010}
N.~Strohmaier{ }{\it et al.}, Phys.\ Rev.\ Lett.\ \textbf{104}, 080401 (2010).

\end{thebibliography}
\end{document}